\documentclass[10pt,aps,prx,twocolumn,superscriptaddress,floatfix]{revtex4-2}

\pdfoutput=1
\usepackage[utf8]{inputenc}

\usepackage{amsmath}
\usepackage{amssymb}
\usepackage{braket}

\usepackage{graphicx}
\usepackage{color}
\usepackage[english]{babel}
\usepackage{microtype}

\usepackage{hyperref}
\hypersetup{
	pdftoolbar=true,
	pdfmenubar=true,
	pdffitwindow=false,
	pdfstartview={FitH},
	pdftitle={Planckian Metal at a Doping-Induced Quantum Critical Point},
	pdfauthor={Dumitrescu, Wentzell, Georges, Parcollet},
	pdfsubject={},
	pdfcreator={},
	pdfproducer={},
	pdfnewwindow=true,
	colorlinks=true,
	linkcolor=black,
	citecolor=blue,
	filecolor=magenta,
	urlcolor=blue
}

\newcommand{\header}[1]{\vspace{1em}\noindent\textbf{#1 --}}

\renewcommand{\vec}{\boldsymbol}

\renewcommand{\Re}{\textrm{Re}}
\renewcommand{\Im}{\textrm{Im}}

\newcommand{\ud}{\mathrm{d}}
\newcommand{\vare}{\varepsilon}

\newcommand{\phdag}{{\phantom{\dagger}}}
\newcommand{\doping}{p}
\newcommand{\dopstar}{p^\star}
\newcommand{\taustar}{\tau^\star}
\newcommand{\iomn}{i\omega_n}
\newcommand{\Sigmat}{\overline{\Sigma}}
\newcommand{\taut}{\overline{\tau}}
\newcommand{\compress}{\chi_e}

\newcommand{\eqnref}[1]{Eq.~\eqref{#1}}

\newcommand{\CCQ}{Center for Computational Quantum Physics, Flatiron Institute, 162 5th Avenue, New York, NY 10010, USA}

\begin{document}
\title{Planckian Metal at a Doping-Induced Quantum Critical Point}

\author{Philipp T.~Dumitrescu}
\email{pdumitrescu@flatironinstitute.org}
\affiliation{\CCQ}

\author{Nils Wentzell}
\affiliation{\CCQ}

\author{Antoine Georges}
\affiliation{\CCQ}
\affiliation{Collège de France, 11 place Marcelin Berthelot, 75005 Paris, France}
\affiliation{Centre de Physique Théorique, Ecole Polytechnique, CNRS, 91128 Palaiseau Cedex, France}
\affiliation{Department of Quantum Matter Physics, University of Geneva, 24 Quai Ernest-Ansermet, 1211 Geneva 4, Switzerland}

\author{Olivier Parcollet}
\affiliation{\CCQ}
\affiliation{Universit\'e Paris-Saclay, CNRS, CEA, Institut de Physique Th\'eorique, 91191, Gif-sur-Yvette, France}

\date{\today}

\begin{abstract}
We numerically study a model of interacting spin-$1/2$ electrons with random
exchange coupling on a fully connected lattice. This model hosts a quantum
critical point separating two distinct metallic phases as a function of doping:
a Fermi liquid phase with a large Fermi surface volume and a low-doping phase
with local moments ordering into a spin-glass. We show that this quantum
critical point has non-Fermi liquid properties characterized by $T$-linear
Planckian behavior, $\omega/T$ scaling and slow spin dynamics of the
Sachdev-Ye-Kitaev (SYK) type. The $\omega/T$ scaling function associated with
the electronic self-energy is found to have an intrinsic particle-hole
asymmetry, a hallmark of a `skewed' non-Fermi liquid.
\end{abstract}

\maketitle

The normal-state properties of hole-doped cuprates are fundamentally different
on the two sides of the critical doping $\doping=\dopstar$ at which the
pseudogap opens. For $\doping>\dopstar$ the Fermi surface (FS) is large and
consistent with band structure predictions~\cite{Hussey_2003,Proust_2019}. In
contrast, for $\doping<\dopstar$ there is clear experimental evidence that a
transformation to a `small' FS takes
place~\cite{Doiron_Leyraud_2007,Proust_2019,Fang_2020}. The vicinity of
$\dopstar$ hosts a `strange metal' in which resistivity is linear in temperature
$T$ down to low-$T$ (for reviews, see
\cite{Hussey_2008,Taillefer_2010,Varma_2020}). Hallmarks of quantum
criticality~\cite{Michon_2019} have been reported in this regime including
$\omega/T$ scaling in spectroscopy experiments~\cite{Marel2003,Reber_2019}. The
nature of the $\doping<\dopstar$ phase and that of the strange metal are
outstanding fundamental questions.

Microscopic models that exhibit such a doping-induced quantum critical point
(QCP) and can also be investigated in a controlled manner are rare. In early
pioneering work, Sachdev and Ye~\cite{SY} showed that the random-bond fully
connected quantum Heisenberg model hosts a spin-liquid phase when solved for
$SU(M)$ spins in the large-$M$ limit. Remarkably, the local spin dynamics in
this phase has the characteristic frequency dependence of a marginal Fermi
liquid~\cite{Varma1989,VarmaIOP,Varma_2020} and obeys $\omega/T$ scaling as a
consequence of conformal invariance~\cite{parcollet1999}. A generalization to a
$t$-$J$ model including itinerant charge carriers was introduced by two of the
present authors~\cite{parcollet1999} (see also
\cite{Florens_2013,balents2017,senthil2018,Patel2019}), who found that in the
large-$M$ limit there is a QCP at zero doping. The doped metal was found to be a
Fermi liquid (FL) at low-$T$, with a higher-$T$ quantum-critical regime
corresponding to a `bad
metal'~\cite{emery_kivelson_prl_1995,calandra_gunnarsson_prb_2002,Hussey_badmetal_2004,Deng2013,Hartnoll_2014}
with $T$-linear resistivity larger than the Mott-Ioffe-Regel value.

Triggered by widespread interest in the broader Sachdev-Ye-Kitaev (SYK)
framework and duality to quantum
gravity~\cite{kitaev_talk,Sachdev_2015,Maldacena2016}, this line of research has
recently been substantially
revived~\cite{Cha2020,Joshi_2020,Tikhanovskaya_I_2020,Tikhanovskaya_II_2020,Shackleton_2020}.
The realistic case of spin-$1/2$ $SU(2)$ electrons is much richer than the
large-$M$ limit considered in early works~(see however
\cite{Tikhanovskaya_I_2020,Tikhanovskaya_II_2020}).  Unlike the large-$M$ model,
the undoped $SU(2)$ insulator has a spin glass ground-state and a finite
ordering temperature~\cite{Bray1980,Grempel1998,Arrachea2002}.
At half-filling, the QCP associated with the melting of this spin-glass phase by
charge fluctuations was recently studied in Ref.~\cite{Cha2020}. It has been
shown that the $SU(2)$ doped model hosts a QCP at a finite critical doping
$\doping=\doping_c$~\cite{Otsuki2013,Joshi_2020,Shackleton_2020}. Understanding
the properties of this QCP and whether it shares some of the properties of
cuprate phenomenology despite the highly simplified character of the model is a
fundamental and fascinating question which is currently being actively
investigated~\cite{Shackleton_2020,Tikhanovskaya_I_2020,Tikhanovskaya_II_2020}.

\begin{figure}[b] 
    \centering
    \includegraphics{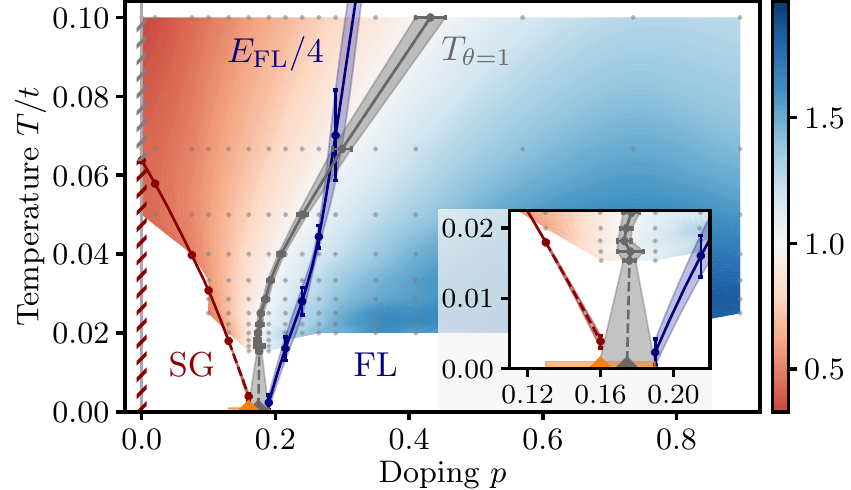}
    \caption{
    \label{fig:phase-diagram}  
    Phase diagram in temperature~$T$ and doping~$\doping$. A quantum critical
    point (QCP; orange dot) separates a spin-glass ordered phase (SG) at low
    $\doping$ from a Fermi liquid (FL) at large $\doping$. Both the SG
    transition temperature (dark red) and the characteristic FL scale
    $E_\mathrm{FL}/ 4$ (dark blue) collapse at or near the QCP. The background
    color corresponds to the power~$\theta$ of the long-time spin correlation;
    the dark grey data and line demarcate $\theta=1$. Pale grey dots indicate
    parameter values at which calculations were performed. Dashed lines are
    extrapolations outside the available temperature range. Inset:~close-up of
    the QCP.}
\end{figure}

In this article, we show that the quantum critical regime associated with the
finite doping QCP hosts a strange metal in which the lifetime of excitations
shows `Planckian' behavior $1/\taustar = c\, k_BT/\hbar$, with $c$ of order
unity~\cite{Zaanen04,Homes_2004,Bruin_2013,Legros19,Grissonnanche_2020,Varma_2020,Sadovskii_2020}.
Furthermore, we demonstrate $\omega/T$ scaling of both the local spin dynamics,
which is found to be of SYK type as in the undoped case~\cite{Cha2020}, and of
the single-particle properties such as the frequency-dependent scattering rate.
In the accessible range of temperature, the latter is found to display a
particle-hole asymmetry which also scales with $\omega/T$ (`skewed' non-Fermi
liquid~\cite{georges2021skewed}).
We establish the phase diagram displayed in Fig.~\ref{fig:phase-diagram}. At
high doping and low-$T$, the metallic phase is a FL, with a crossover into
Planckian behavior in the quantum critical regime. At the QCP the volume of the
FS changes abruptly and the system transitions from a FL to another metallic
state which is unstable to spin-glass ordering below the indicated critical
temperature.
These results are established using the extended dynamical mean-field theory
(EDMFT)
framework~\cite{Sengupta95,Si_1996,smith_si_prb_2000,Chitra2000,Georges1996} and
a quantum Monte Carlo algorithm~\cite{Rubtsov2005,SM}, allowing us to study
disorder-averaged quantities directly in the thermodynamic limit. In this
regard, our study provides a complementary perspective on the quantum critical
regime to the recent parallel work by Shackleton et.~al.~\cite {Shackleton_2020}
which uses exact diagonalization of finite-size systems for a fixed
configuration of disorder.

\header{Model}
We consider spin-1/2 electrons on a fully connected lattice governed by the
Hamiltonian
\begin{equation}
H = H_0+
\sum_{i<j} {J_{ij}\vec{S}_i \cdot \vec{S}_j} + U \sum_{i} n_{i \uparrow} n_{i \downarrow},
\label{eq:HamFull}
\end{equation}
where $H_0 =  - \sum_{ij, \sigma} (t_{ij} + \mu \delta_{ij}) c^{\dag}_{i\sigma}
c^{\phdag}_{j\sigma} $. Here $\sigma = \uparrow, \downarrow$ is the spin-state,
$\mu$ is the chemical potential, $n_{i\sigma} =
c^{\dag}_{i\sigma}c^{\phdag}_{i\sigma}$ is the occupation number on site $i$
with spin-state $\sigma$, and $\vec{S}_i =  \tfrac{1}{2} c^{\dag}_{i\sigma'}
\vec{\sigma}_{\sigma'\sigma} c^{\phdag}_{i\sigma}$ is the spin operator on site $i$. 
The hopping amplitudes $t_{ij}$ are complex with $t_{ij} = t^*_{ji}$, and the
exchange coupling strengths $J_{ij}$ are real. They are drawn according to
independent Gaussian distributions with zero means $\overline{t_{ij}}=
\overline{J_{ij}}= 0$ and variances $\overline{ \vert t_{ij} \vert^2} = t^2 /
\mathcal{N}$, $\overline{ \vert J_{ij} \vert^2} = J^2/\mathcal{N}$.
Here, $\mathcal{N}$ is the number of lattice sites and we work directly in the
thermodynamic limit $\mathcal{N}\to\infty$,  keeping $t,J,U$ finite
\cite{noteph}. We
 use the replica method to deal with the quenched disorder and we restrict
 ourselves to replica diagonal paramagnetic solutions without spin-symmetry
 breaking.

The local electron and spin correlation functions of \eqnref{eq:HamFull} are
obtained by solving an auxiliary quantum `impurity' model (EDMFT
equations)~\cite{Sengupta95,Si_1996,smith_si_prb_2000,Chitra2000,Georges1996,parcollet1999,Cha2020}
\begin{multline}
 \label{eq:action} S_{\mathrm{eff}} = \int \!\! \ud\tau \left[ \sum_{\sigma}
 c^\dag_{\tau,\sigma} \left[ \partial_\tau - \mu \right] c^\phdag_{\tau,\sigma} 
 + U n_{\tau,\uparrow} n_{\tau,\downarrow}\right] + \\
  \int \!\! \ud\tau' \ud\tau \! \left[ \Delta(\tau' - \tau)
  c^\dag_{\tau',\sigma}c^\phdag_{\tau,\sigma} - \frac{J^2Q(\tau' - \tau)}{2} 
  \vec{S}_{\tau'} \cdot \vec{S}_{\tau} \right]\!\!
 \end{multline}
subject to the self-consistency conditions $\Delta(\tau) = t^2 G(\tau)$ and
$Q(\tau - \tau') = \tfrac{1}{3} \langle \vec{S}(\tau) \cdot \vec{S}(\tau')
\rangle$. Here we denote imaginary time by $\tau\in[0,\beta]$ and inverse
temperature by $\beta=1/T$ ($k_B = 1$ unless otherwise noted). The
self-consistency conditions relate the fermionic hybridization bath $\Delta$ to
the local fermionic Green function $G(\tau) = -\langle T_\tau c(\tau)
c^\dag(0)\rangle$ and the retarded spin-spin interaction $Q$ to the local spin
correlation function. The electronic self-energy of the impurity model $\Sigma(i
\omega_n)  = i\omega_n + \mu - \Delta(i \omega_n) - G^{-1}(i\omega_n)$, where
$\omega_n=(2n+1)\pi/\beta$ 
are fermionic Matsubara frequencies, coincides with the self-energy of model
\eqnref{eq:HamFull}. Therefore, we can reconstruct the Green function of the
lattice model $G_\mathrm{latt}^{-1}(\vare,\iomn)=\iomn+\mu-\vare-\Sigma(\iomn)$,
where $\vare$ is an eigenvalue of the random matrix $t_{ij}$ (see
App.~\ref{app:lattice-gf}). In the thermodynamic limit, these eigenvalues
$\vare$ are distributed according to the Wigner semi-circle law and the local
$G$ and $\Sigma$ are self-averaging in the paramagnetic
phase~\cite{Georges1996,Shackleton_2020,SM}.
The impurity model is solved using a quantum Monte Carlo interaction-expansion
impurity solver in continuous imaginary time (CT-INT)~\cite{Rubtsov2005,
Cha2020}, for details see App.~\ref{app:numerics}. Throughout, we set the interaction
strengths $J = 0.5 t$ and $U = 4t$ -- sufficiently strong to realize the finite
doping QCP \cite{Cha2020}, while still amenable to CT-INT simulations.


\begin{figure*}[ht] 
    \centering
    \includegraphics{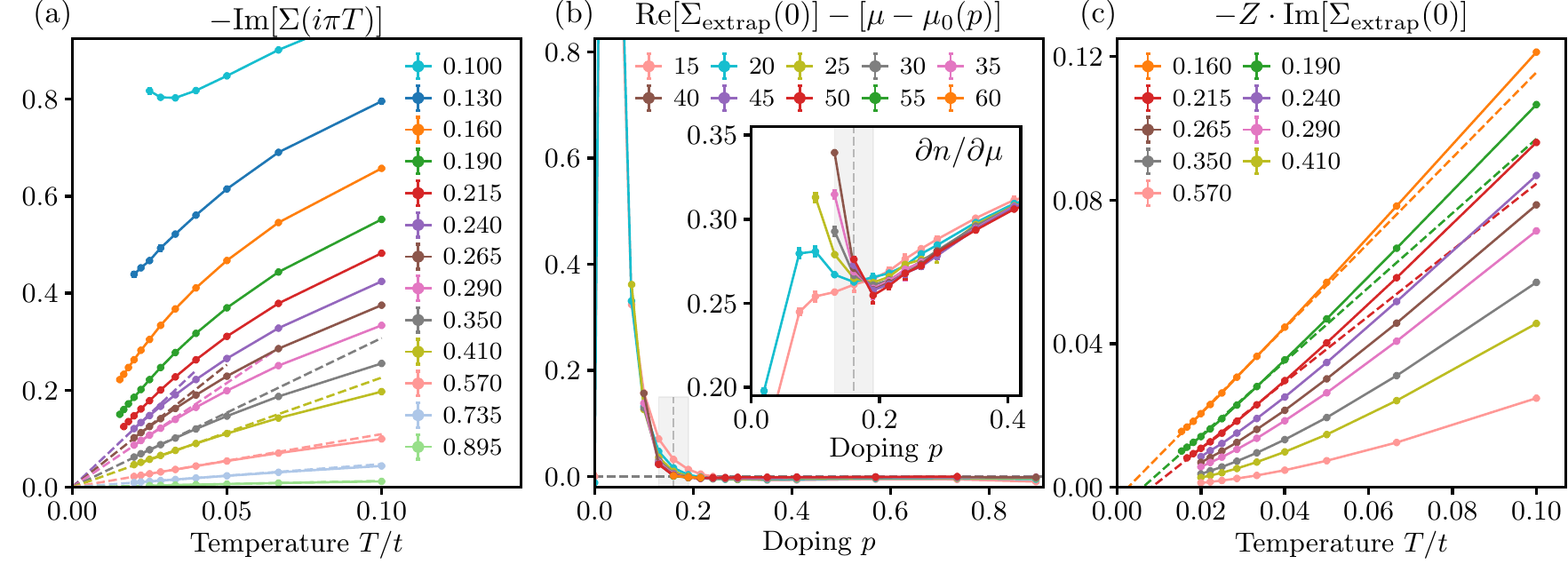}
    \caption{
    \label{fig:electron_sigma} 
    (a) Imaginary component of the self-energy  at the first Matsubara
    frequency $\Im[\Sigma(i \pi T)]$, for various doping levels $p$ from $0.1$
    to $0.895$. The dashed lines emphasize the low $T$-linear regime which
    characterizes the Fermi liquid.
    (b) Violation of the Luttinger theorem at the QCP. The quantity $\Re
    \Sigma_{\mathrm{extrap}}(0) - [\mu - \mu_0(\doping)]$ vs $\doping$ for
    various temperatures $15 \leq \beta \leq 60$ -- it vanishes at low $T$ in
    the high doping Fermi liquid phase. The doping $\doping_c$ where it
    deviates from zero  defines the QCP (vertical dashed line).
    Inset: Electronic compressibility $\compress = \partial n / \partial \mu$.
    (c) Inverse lifetime $1/\taustar = - Z\Im[\Sigma_{\mathrm{extrap}}(0)]$
   (see App.~\ref{app:extrap-imSigma}). It has a Fermi liquid $T^2$
    behavior at
    high doping. Close to the QCP, $p_c\approx 0.16-0.19$, it becomes linear in
    $T$ at low temperatures (dashed lines). }
\end{figure*}


\header{Fermi Liquid}
Let us start from the large doping regime. In a FL, $\Sigma(\iomn)$ at low
$\omega_n$ and $T$ takes the form
\begin{equation}
\label{eq:sigma_fl}
\Im \Sigma(i\omega_n) = \left(1 - \frac{1}{Z} \right) \omega_n + \frac{ \omega_n^2 - (\pi T)^2 }{E_{\mathrm{FL}}}+O(T^3),
\end{equation}
where $Z$ is the quasiparticle weight and $E_{\mathrm{FL}}$ is a characteristic
FL scale~(see App.~\ref{app:lattice-gf}).
Thus, a hallmark of FL behavior is that $\Sigma(i\omega_n)$ at the first
Matsubara frequency depends {\it linearly} on $T$: $\Im \Sigma(i \pi T) \sim T +
O(T^3)$~\cite{Chubukov_2012}.
As shown in Fig.~\ref{fig:electron_sigma}a, this $T$-linear behavior holds up to
a FL crossover-scale $T_\mathrm{FL}$, which is large at large doping and
decreases with $\doping$. At $\doping \leq 0.16$, $\Im \Sigma(i \pi T)$
extrapolates to a finite value at $T=0$, signaling a breakdown of FL behavior.
We can obtain $E_{\mathrm{FL}}$ and $Z$ from a direct fit of $\Im\Sigma(\iomn)$
to \eqnref{eq:sigma_fl}. These quantities, together with $T_{\mathrm{FL}}$,
collapse at the QCP (see App.~\ref{app:supp-data}); see
Fig.~\ref{fig:phase-diagram}.

A sharp signature of the collapse of the FL at the QCP is the sudden violation
of the Luttinger theorem~\cite{Otsuki2013}, which constrains the volume of the
Fermi surface. From the expression of the lattice Green function
$G_\mathrm{latt}$, the interacting FS is defined by
$\vare=\mu-\Re\Sigma_{\mathrm{extrap}}(0)$, where $\Re
\Sigma_{\mathrm{extrap}}(0)$ is the real part of $\Sigma$ extrapolated to zero
frequency~(see App.~\ref{app:supp-data}). Comparing the non-interacting and
interacting Fermi surface, Luttinger's theorem states that $\lim_{T\rightarrow
0}\Delta\vare_F(T) = 0$, where we define $\Delta\vare_F(T)=\Re
\Sigma_{\mathrm{extrap}}(0) - [\mu - \mu_0(\doping)]$, with $\mu_0(p)$ the
chemical potential of the non-interacting system $U=J=0$ at the same doping.
Figure~\ref{fig:electron_sigma}b shows $\Delta\vare_F(T)$ versus doping for
various temperatures. While for large doping levels, it converges to zero at low
temperature as expected for a FL, it clearly does not at low doping. The doping
where the deviation onsets, defines the critical point $\doping_c = 0.16 \pm
0.03$. For $\doping < \doping_c$ the FL is destroyed and replaced by another
metallic phase discussed below.


\header{Planckian behavior.} 
We now discuss the QCP, approaching it from the high-doping side.
Figure~\ref{fig:electron_sigma}c shows $1/\taustar=-Z\Im
\Sigma_{\mathrm{extrap}}(0)$, which is the width of the spectral function
$A(\vare,\omega)=-\mathrm{Im}G(\vare,\omega+i0^+)/\pi$. In the FL regime
$1/\taustar \propto T^2$ and can be interpreted as the inverse of the
quasiparticle lifetime. Close to the QCP, FL behavior breaks down and we find a
clear `Planckian'
behavior~\cite{Zaanen04,Homes_2004,Bruin_2013,Legros19,Grissonnanche_2020,Varma_2020,Sadovskii_2020}
down to low-$T$
\begin{equation}
    \frac{1}{\taustar}\simeq c\,\frac{k_B T}{\hbar},
    \label{eq:planckian}
\end{equation}
restoring fundamental constants. Here $c$ is a coefficient of order unity;
$c=1.0 \pm 0.1$ for $\doping=0.19$.
Furthermore, the transport scattering rate $1 / \tau_{\mathrm{tr}} =
-2\Im\Sigma_{\mathrm{extrap}}(0)$ is also approximately $T$-linear in this
regime~(see App.~\ref{app:supp-data}). Since the self-energy $\Sigma(\omega)$ is
strictly local, the electrical resistivity $\rho$ defined via the Kubo
formula~(see App.~\ref{app:conductivity}) is determined by
$1/\tau_{\mathrm{tr}}$. This implies that $\rho$ has an approximately $T$-linear
dependence. We emphasize that the resistivity is smaller than the
Mott-Ioffe-Regel value at low-$T$, in contrast to `bad metal' behavior.
When viewed in terms of Einstein-Sutherland relation
$1/\rho=D\,\chi_e$~\cite{calandra_gunnarsson_prb_2002,Hartnoll_2014,Perepelitsky2016,Park_Diffusivity_2020},
the $T$-linearity of $\rho$ stems from the diffusion constant $D\propto 1/T$,
rather than from the compressibility $\chi_e=\partial n/\partial\mu$, which has
little $T$-dependence at the QCP (Fig.~\ref{fig:electron_sigma}b inset).

\begin{figure*}[ht] 
    \centering
    \includegraphics{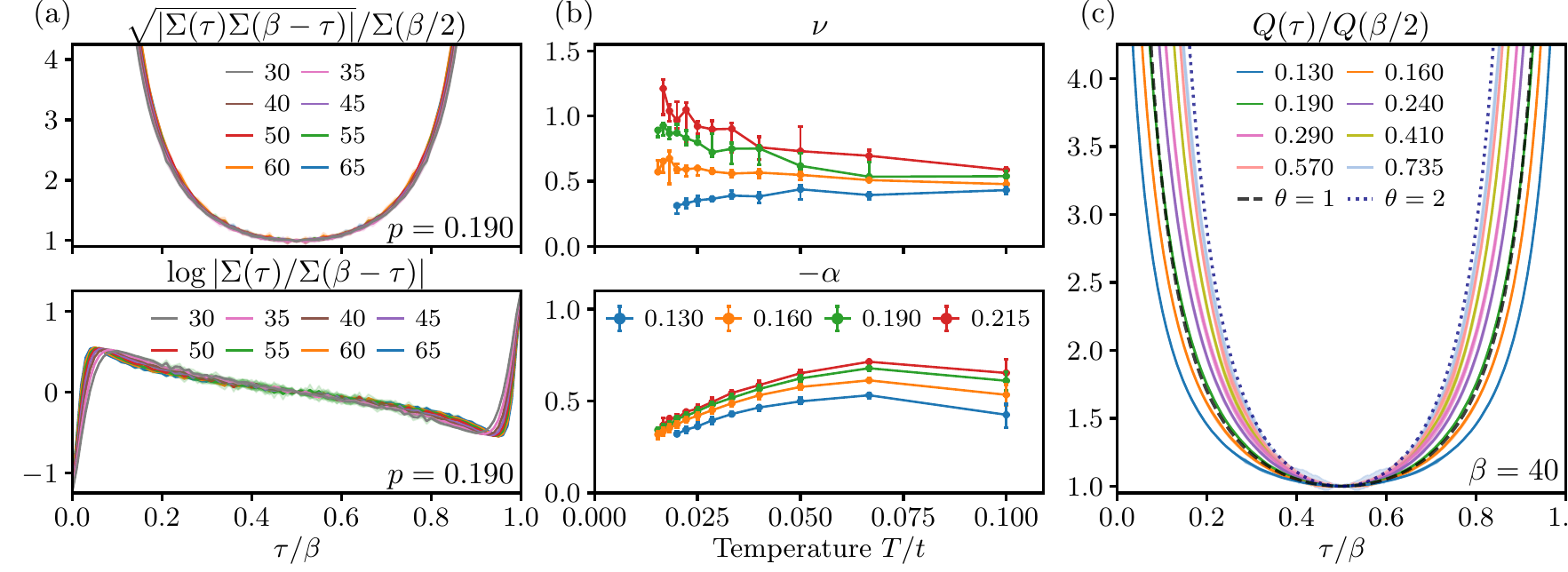}
    \caption{
    \label{fig:scaling_powers} 
    (a) Symmetric
    (top) and anti-symmetric (bottom) scaling forms of the electronic self-energy
    in imaginary time $\Sigma(\tau)$ vs $\tau/\beta$. Doping is fixed $\doping = 0.19$ to be near
    the QCP and we show various $\beta$. 
    (b) Non-Fermi liquid exponent $\nu$ (top) and asymmetry parameter $\alpha$ (bottom), 
    extracted by fitting the scaling form in (a) to \eqnref{eq:sigmaScaling}, 
    for various doping levels $\doping$ close to the QCP~\cite{noteph}. 
    (c) Normalized spin-spin correlation $Q(\tau) / Q(\beta / 2)$ at a fixed $\beta = 40$ and varying $\doping$. The long imaginary times scaling form
	$Q(\tau) \sim 1 / [\sin(\pi \tau / \beta)]^\theta$ is shown with $\theta = 2$ (Fermi liquid, dotted line)  and $\theta = 1$ (SYK, dashed line).
     }
\end{figure*}
\header{Quantum criticality: skewed non-Fermi liquid and $\omega/T$ scaling.} 
We now show that our data support $\omega/T$ scaling of the self-energy near
the QCP. In real-frequencies, we expect a scaling form $-\Im\Sigma(\omega+i0^+)
\propto T^\nu \sigma(\omega/T)$ with $\nu$ an exponent ($\nu=2$ for a Fermi
liquid). This translates in imaginary time to
$\Sigma(\tau)/\Sigma(\beta/2)=\Sigmat(\taut)$ with $\taut=\tau/\beta$. In order
to test this scaling form and identify the scaling function $\Sigmat$, we plot
in Fig.~\ref{fig:scaling_powers}a
$\sqrt{\vert\Sigma(\tau)\Sigma(\beta-\tau)\vert}/\Sigma(\beta/2)$ and
$\log\vert \Sigma(\tau) / \Sigma(\beta-\tau)\vert$ for several $\beta$ and a
fixed $\doping = 0.19$ close to $\doping_c$. This allows to address separately
the symmetric (even) and antisymmetric (odd) components of $\Sigma$ under
particle-hole symmetry $\tau\rightarrow \beta-\tau$ ($\omega\rightarrow
-\omega$). Within the range of temperature accessible to our algorithm we
obtain a good scaling collapse of the data in the long-time limit around $\taut
= 1/2$~(see App.~\ref{app:supp-data}). The scaling function agrees well with the conformally
invariant ansatz:
\begin{equation}
\label{eq:sigmaScaling}
\Sigmat(\taut)\,=\,\,
e^{\alpha(\taut - 1/2)}\,\left[\frac{1}{\sin\pi\taut}\right]^{\nu + 1},\qquad \taut=\frac{\tau}{\beta}.
\end{equation}
Figure~\ref{fig:scaling_powers}b displays the values of $\nu$ and $\alpha$
obtained from a fit of the data in Fig.~\ref{fig:scaling_powers}a. We note that
$\nu$ varies substantially close to the QCP. The marginal Fermi liquid value
$\nu=1$~\cite{Varma1989,VarmaIOP,Varma_2020} and the $SU(M\rightarrow\infty)$
model value $\nu = 1/2$~\cite{SY,parcollet1999} are both consistent with our
data in the low-$T$ limit, but lie at opposite ends of our extrapolated range.
We also note that our observed $T$-linear behavior of $1/\tau_\mathrm{tr}$ has
to arise out of a combination of the finite temperature dependence of the
effective $\nu,\alpha$ and prefactor of $\Sigma(\tau)$ (see
App.~\ref{app:scaling-fn}).

Remarkably, Fig.~\ref{fig:scaling_powers}b shows that at finite-$T$ in the
quantum critical region, our model behaves as a `skewed' non-Fermi liquid, with
an $\omega/T$ scaling function $\sigma$ displaying an intrinsic particle-hole
asymmetry. The latter is encoded in the spectral asymmetry parameter (skew)
$\alpha$ of \eqnref{eq:sigmaScaling} (see App.~\ref{app:scaling-fn}), which
takes rather large values at finite $T$. Whether this asymmetry persists down to
zero temperature at the QCP (i.e.~$\alpha$ remains finite at $T=0$) is an
interesting open question. Recently, such a particle-hole asymmetry in skewed
Planckian metals attracted strong interest, from both
theory~\cite{georges2021skewed} and experiments~\cite{Gourgout2021}, as a
possible explanation of a puzzle regarding the sign and $T$-dependence of the
Seebeck coefficient in cuprate superconductors. Measurements of the thermopower
of twisted bilayer graphene~\cite{Ghawri2020} indicate possible relevance to
other materials as well. We also note that the skew is itself of basic
theoretical interest; in the large-$M$ limit, there is a fundamental
relationship between it and finite entropy density at the
QCP~\cite{parcollet_kondo_prb_1998}.


\header{Quantum criticality: SYK spin dynamics} 
Figure~\ref{fig:scaling_powers}c shows the local spin-spin correlation function
$Q(\tau)$ at $\beta=1/40$, for several doping levels. In imaginary time, the
universal long time scaling is around $\tau = \beta/2$. By comparison to the
conformal scaling function~\cite{parcollet1999,SM} $Q(\tau)/Q(\beta/2) \sim
1/\left[\sin(\pi\tau/\beta)\right]^\theta$, we see that at the QCP the spin
dynamics slows down from the long-time behaviour $\sim 1/\tau^2$ characteristic
of a Fermi liquid ($\theta=2$) to the SYK dynamics~\cite{SY} $\sim 1/\tau$
($\theta=1$).
We fit the critical exponent $\theta$ for all $T$ and $\doping$, and display it
as the background color in Fig.~\ref{fig:phase-diagram}. This allows us to
locate the temperature scale $T_{\theta=1}$ at which $\theta = 1$
(Fig.~\ref{fig:phase-diagram}, dark grey data), which extrapolates to the
quantum critical point $\doping_c$ at low-temperatures within error bars. This
critical SYK scaling was also found in a renormalization group
analysis~\cite{Sengupta2000,Vojta2000,Joshi_2020} and for the undoped model
\cite{Cha2020}.

\begin{figure}[t] 
    \centering
    \includegraphics{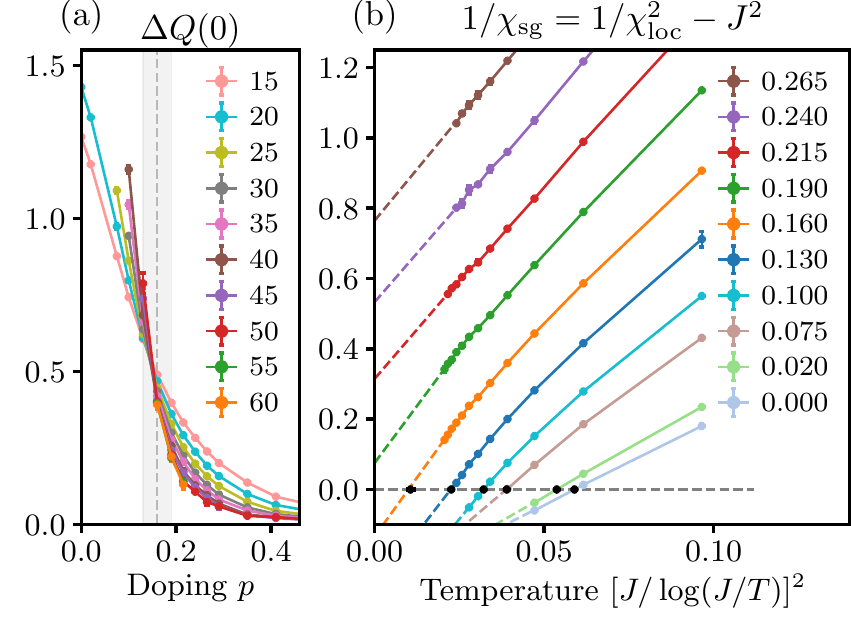}
    \caption{
    \label{fig:chi} 
    (a) Estimate of local moment component of the spin response $\Delta Q(0)$
    (see text). The dashed line marks the QCP $\doping_c$. 
    (b) Inverse of the spin-glass susceptibility $\chi_{\mathrm{sg}}$ vs
    $\bigl[J/\log(J/T) \bigr]^2$. Black dots mark the extrapolated transition
    temperature. Dashed lines are linear extrapolations.
    }
\end{figure}

\header{Low doping metal and spin-glass}
The critical doping $\doping_c$ separates a FL at $\doping>\doping_c$ from a 
phase of a different nature for $\doping<\doping_c$. 
This phase is also metallic -- as seen numerically from $G(i\omega_n)$ at low
frequency~(see App.~\ref{app:supp-data}) -- but it has emerging local moments~\cite{Otsuki2013}.
This is demonstrated by Fig.~\ref{fig:chi}a, which shows $\Delta Q(0) =
\chi_{\mathrm{loc}}  - Q_{\mathrm{extrap}} (0)$, where $\chi_{\mathrm{loc}} =
Q(i\nu_0)$ is the static local spin susceptibility at the bosonic Matsubara
frequency $\nu_0=0$ and $Q_{\mathrm{extrap}}(0)$ is the extrapolated value of
$Q(i\nu_n)$ for $\nu_n\rightarrow 0$.
A local moment $m=\langle S^z\rangle$ is associated with a plateau in the
spin-spin correlation $Q(\tau)\sim m^2$ at long time. Hence, its Fourier
transform is $Q(i\nu_n)= \beta m^2\delta_{n,0}+Q_{\mathrm{reg}}(i\nu_n)$, where
$Q_{\mathrm{reg}}$ is a regular (decaying) function, so that $\Delta
Q(0)\propto \beta m^2$.
For $p>p_c$,  $\Delta Q(0)$ decreases to zero upon cooling, while it grows for
$p<p_c$, a clear signature of local moments at low $T$.
The presence of local moments for $\doping<\doping_c$ also explains the
distinctive change of behavior of the compressibility $\compress$ through the
critical point  (Fig.~\ref{fig:electron_sigma}b). The $T$-dependence of
$\compress$ is related to the entropy per site $s$ by the Maxwell relation
$\partial^2 s/\partial n^2|_{T}=\compress^{-2}\,\partial\compress/\partial
T|_{n}$. In a local moment phase, we expect $s$ to be finite at $T=0$ (and
$\partial^2 s/\partial n^2<0$; see App.~\ref{app:entropy}), while in the FL phase the entropy
vanishes as $s\propto \gamma T$. Hence, at fixed $T$, $s(n)$ must have an
inflection point $\partial^2 s/\partial n^2 =0$, implying that at this density
the compressibility must be independent of $T$. This is indeed observed for
$\doping\simeq \doping_c$ on Fig.~\ref{fig:electron_sigma}b (inset).

This local moment metallic solution of the paramagnetic EDMFT equations is
unstable to spin glass ordering below the critical temperature depicted on
Fig~\ref{fig:phase-diagram}. The spin-glass susceptibility is given
by~\cite{GPS00,GPS01} $\chi_{\mathrm{SG}} \propto \chi_{\mathrm{loc}}^2 / (1 -
J^2 \chi_{\mathrm{loc}}^2 )$. Its $T$-dependence is displayed on
Figure~\ref{fig:chi}b, where we plot $\chi_{\mathrm{SG}}^{-1}$ as a function of
$\bigl( J/\log(J/T) \bigr)^2$. In this representation, we expect a linear
dependence close to the QCP, since we expect from theory that $J
\chi_{\mathrm{loc}} \sim \log (J
/ T)$ at
$p=p_c$ ($T\ll J$ here)~\cite{GPS00}. It is seen that
$\chi_{\mathrm{SG}}$
diverges at a finite spin-glass ordering temperature for $\doping<\doping_c$
(black dots in Fig.~\ref{fig:chi}b).
If the logarithmic form of $\chi_{\mathrm{loc}}$ holds to $T=0$ at the QCP,
then $\chi_\mathrm{SG}$ will diverge at a finite $T$. Therefore, the spin-glass
phase extends to $p\gtrsim p_c$ at very low $T$, although this effect is below
the temperature resolution of the numerical data. Finally, since we restricted
ourselves to paramagnetic (replica diagonal) solutions, we do not describe the
spin-glass phase itself. We expect the phase to be metallic at non-zero doping.

\header{Conclusion} 
The QCP analyzed in this paper separates two distinct metallic phases: a Fermi
liquid with a FS volume consistent with Luttinger theorem for
$\doping>\doping_c$ and a phase with local moments and a spin-glass instability
for $\doping<\doping_c$.
While being consistent with previous work on this
model~\cite{Otsuki2013,Joshi_2020,Shackleton_2020}, our study offers new
insights into its physical properties.
We have shown that a  skewed non-Fermi liquid emerges, characterized by
$T$-linear `Planckian' behavior, slow SYK spin dynamics, $\omega/T$ scaling and
a low-energy particle-hole asymmetry also scaling as $\omega/T$ (`skew').
It is striking that, despite the extreme simplification of the fully connected
$SU(2)$ random $t$-$J$-$U$ model studied here, this is reminiscent of some key
aspects of the cuprate
phenomenology~\cite{Marel2003,Varma_2020,Taillefer_2010,Proust_2019}. We note
that recent nuclear magnetic resonance and ultrasound measurements have revealed
that, remarkably, the spin-glass phase of La$_{2-x}$Sr$_x$CuO$_4$ extends up to
$\doping=\doping^\star$ in high magnetic fields~\cite{frachet2020hidden}.
We also note that $\omega / T$ scaling of the self-energy has also been reported
in ARPES experiments \cite{Reber_2019}.
Our results may be relevant to other materials as well, in which Planckian
behavior and quantum criticality are observed, such as twisted bilayer
graphene~\cite{Park_Diffusivity_2020,Cao_StrangeMagicAngle_2020}.
The `skew' found in our study is of special current
interest~\cite{georges2021skewed} in relation to the unconventional behavior of
the Seebeck coefficient in cuprate superconductors~\cite{Gourgout2021} and
possibly also twisted bilayer graphene~\cite{Ghawri2020}.
%


\vspace{2em}
We thank S.~Sachdev, H.~Shackleton, A.~Wietek as well as P.~Cha, E.-A. Kim and
J.~Mravlje for valuable discussions and collaborations on related work. The
algorithms used in this study were implemented using the TRIQS code
library~\cite{TRIQS2015} and CTINT~\cite{WentzellCTINT}. The Flatiron Institute
is a division of the Simons Foundation.

\nocite{Rubtsov2005, Gull2011Review}
\nocite{WentzellCTINT}
\nocite{Rubtsov2005}
\nocite{Shackleton_2020,Georges1996}
\nocite{parcollet_kondo_prb_1998}
\nocite{Georges1996}
\nocite{balents2017,senthil2018}
\nocite{Georges1996}
\nocite{Sadovskii_2020}
\nocite{Bruin_2013,Legros19,Grissonnanche_2020}
\nocite{Cha2020}

\bibliography{syk-doping-bib}

\begin{thebibliography}{65}
\bibitem [{not()}]{noteph}%
  \BibitemOpen
  \href@noop {} {}\bibinfo {note} {We perform computations for electron doping
  $\doping$; $n = 1+ \doping$, where $n$ is the electronic density. Due to the
  model's symmetry, our results are also true for hole doping, if we change the
  sign of $\Re G(i\omega_n), \Re\Sigma(i\omega_n)$ and particle-hole symmetry
  odd quantities such as $\alpha$, $\Delta \varepsilon_F$.}\BibitemShut {Stop}%
\end{thebibliography}

\onecolumngrid
\clearpage

\renewcommand{\theequation}{S\arabic{equation}}
\renewcommand{\thefigure}{S\arabic{figure}}
\renewcommand{\thetable}{S\arabic{table}}

\setcounter{equation}{0}
\setcounter{figure}{0}
\setcounter{table}{0}

\section*{Appendix}


\subsection{Details of Numerical Simulations} 
\label{app:numerics}

We solve the effective impurity model using the standard dynamical mean field
theory (DMFT) approach. In particular, we iteratively solve \eqnref{eq:action}
and  update the hybridization function $\Delta(\tau) = t^2 G(\tau)$ and the
spin-correlation function $Q(\tau)$ until reaching convergence. At each step,
we use a quantum Monte Carlo impurity solver in continuous imaginary time,
based on perturbative expansion of the partition function in powers of the
fermion interaction strength (CT-INT). We refer to the literature for details
of the algorithms \cite{Rubtsov2005, Gull2011Review}, but note that our
implementation \cite{WentzellCTINT} extends this approach with a fast
computation of $Q(\tau)$ from the three-point vertex function rather than
through an operator insertion measurement.

Since the Monte Carlo sampling is performed in the grand canonical ensemble, we
adjust the chemical potential $\mu$ between DMFT iterations in order to target
a fixed electron density $n$. The iterative procedure converges to a fixed $n$
and $\mu$ alongside the correlation functions.  Each calculation is deemed
converged when the relative change of $\Re G(i \omega_n)$,  $\Im G(i \omega_n)$
and $Q(i \omega_n)$ between two consecutive iterations is smaller than $1\%$.
Note that we apply this convergence only on Fourier modes where there is
appreciable weight ($\vert \Re G(i \omega_n) \vert > 10^{-3}, \vert \Im G(i
\omega_n) \vert > 10^{-3}$, or $Q(i\omega_n) > 10^{-2}$) in order to be above
the level of statistical noise of the Monte Carlo calculation.

To accelerate the DMFT convergence, we seed the initial configuration for a
given $n$ and $T$ from the converged solution at the nearest higher temperature
and same density $n$. Since we are only considering paramagnetic solutions in
\eqnref{eq:action}, this annealing procedure is sufficient and  we do not
consider coexistence or metastability with other orders by construction. Each
parameter runs for at least $10$ iterations until convergence and often more --
typically $\sim 15$ iterations in the Fermi liquid regime and $\gtrsim 25$
iterations in the regime $\doping \leq \doping_c$.
 
Even with these choices, the dominant source of error is still the DMFT
convergence error rather than the statistical Monte Carlo error. The error-bars
throughout the paper are therefore solely based estimates of this systematic
error, obtained by considering the minimum-to-maximum range of values of the
observable in the last five DMFT iterations. 

\subsubsection{Alpha Shift}

The Monte Carlo sampling we perform suffers from a sign problem, which becomes
more pronounced at larger doping. When directly sampling the action
\eqnref{eq:action} with the parameters studied ($U = 4t, J=0.5t$), the sign
problem typically limits calculations to high temperatures $\beta \lesssim 20$
in the Fermi liquid regime. To alleviate the sign problem we use a so-called
$\alpha$-shift. This changes the origin of the perturbative expansion, by
shifting the interaction term of the action by a quadratic term -- the Hubbard
term becomes $U \sum_i (n_{i, \uparrow} - \alpha_{\uparrow}) (n_{i, \downarrow}
- \alpha_{\downarrow})$ -- which is then compensated by a redefinition of the
bare propagator; see e.g.~Ref.~\cite{Rubtsov2005} for details.
For our model, we found it convenient to use an automated optimization
procedure to obtain suitable values for $\alpha_{\uparrow, \downarrow}$. We
perform a single iteration calculation with a reduced number of Monte Carlo
cycles and use the average sign as our optimization function. Since the Monte
Carlo with a low number of measurements has a significant statistical error, we
used a Bayesian optimization based on Gaussian processes, since this can
naturally treat optimization functions with noisy output. Instead of a full 2d
parameter optimization, we found it more efficient to perform 1d optimizations
of $\overline{\alpha} = \tfrac{1}{2}(\alpha_{\uparrow} + \alpha_{\downarrow})$
for increasing values of $\delta = \alpha_{\uparrow} - \alpha_{\downarrow}$
until finding values for which the average sign is $>0.37$.

This procedure works well to ameliorate the sign problem at intermediate
temperatures, but can become insufficient at very low temperatures. In the main
text we limit $\beta \leq 50$ in the Fermi liquid regime (limited primarily by
the sign problem) and $\beta \leq 65$ near the QMC (limited primarily by slow
DMFT convergence).

\subsubsection{Analysis}

For the analysis of several quantities (e.g.~Luttinger's theorem, long-time
spin-spin correlation), we need to perform a numerical extrapolation from finite
Matsubara frequencies to $\omega_n \to 0$ at finite $T$. Since $\omega_n \sim
T$, it is generally challenging to reconstruct the limit $\omega \ll T$ from
imaginary time simulation, unless the full analytic form of the function is
known. In the main text, we use an agnostic local polynomial spline fit using
the first few Matsubara frequencies. An exception is the fermion scattering
rate, where we use a scaling ansatz for a more accurate fit; this is discussed
below.

\subsection{Lattice Green function and Luttinger theorem}
\label{app:lattice-gf}

The local Green function $G_{ii}(i\omega_n)$ and local self-energy
$\Sigma_{ii}(i\omega_n)$ become self-averaging in the thermodynamic limit
$\mathcal{N}\rightarrow\infty$, i.e. do not depend on the site $i$ or on the
random sample $\{t_{ij},J_{ij}\}$ (for details, see
Refs.~\cite{Shackleton_2020,Georges1996}). This can be shown using, for example,
the cavity method and applies to paramagnetic phases but not to the spin-glass
phase.
 
The spectrum of eigenstates of the hopping matrix $t_{ij}$ is distributed
according to a Wigner semi-circle law in the thermodynamic limit. Denoting by
$\ket{\lambda}$ an eigenstate of $t_{ij}$ for a given sample, with eigenvalue
$\vare_\lambda$, the lattice Green function is given by, for
$\mathcal{N}\rightarrow\infty$:
 \begin{equation}
     G_{ij}(i\omega_n) = \sum_\lambda \braket{i|\lambda} G(i\omega_n,\vare_\lambda) 
     \braket{\lambda|j},
 \end{equation}
where
 \begin{equation}
     G(i\omega_n,\vare) = \frac{1}{i\omega_n+\mu-\vare-\Sigma(i\omega_n)}.
 \end{equation}
In the Fermi Liquid regime, the self-energy at low temperature is
\begin{equation}
\Sigma(i\omega_n) = \Re\Sigma(i\omega_n) +i \left[\left(1 - \frac{1}{Z} \right)
\omega_n + \frac{ \omega_n^2 - (\pi T)^2 }{E_{\mathrm{FL}}}\right],
\end{equation}
where $Z$ is the quasiparticle weight and $E_{\mathrm{FL}}$ is a characteristic
Fermi liquid scale. Hence,
 \begin{equation}
     G(i\omega_n,\vare) =
     \frac{Z}{i\omega_n+Z\left[\mu-\vare-\Re\Sigma(i\omega_n)\right] - i Z({
     \omega_n^2 - (\pi T)^2 })/{E_{\mathrm{FL}}}}.
 \end{equation}
The Fermi surface is thus located at the single particle energy
 \begin{equation}
     \vare = \mu-\Re\Sigma(0,T=0) \equiv \vare_F.
 \end{equation}
Luttinger's theorem states that, for a fixed particle density $n$,
$\mu(T=0,n)-\Re\Sigma(0,T=0,n)=\mu_0(T=0,n)$. Here $\mu_0$ is the
non-interacting value of the chemical potential at the same density (i.e with
$U=J=0$).
 
\subsection{Skewed $\omega/T$ scaling functions}
\label{app:scaling-fn}
 
The conformally invariant scaling form for the self-energy in imaginary time
reads~\cite{parcollet_kondo_prb_1998}
 \begin{equation}
\frac{\Sigma(\tau)}{\Sigma(\beta/2)}\,=\,
e^{\alpha(\taut-1/2)}\,\left(\frac{1}{\sin \pi\taut} \right)^{1+\nu}, \qquad
\taut=\frac{\tau}{\beta}.\label{eq:scaling_tau}
\end{equation}
This function has the following spectral representation
\begin{equation}
e^{\alpha(\taut-1/2)}\,\left(\frac{1}{\sin \pi\taut} \right)^{1+\nu}\,=
\int_{-\infty}^{+\infty}
dx\,\frac{e^{-x\taut}}{1+e^{-x}}\,\widetilde{\sigma}_{\alpha,\nu}(x)
\end{equation}
with $x=\omega/T$ and 
 \begin{equation}
 \label{eq:sigma_scaling_x}
   \widetilde{\sigma}_{\alpha,\nu}(x)\,=\,
  \frac{2^\nu}{\pi^2\Gamma(1+\nu)}\cosh\frac{x}{2}\, 
   \bigg|\Gamma \left[\frac{1+\nu}{2}+ i \frac{x+\alpha}{2\pi}\right]\bigg|^2.
\end{equation}
In the Planckian case $\nu=1$ this expression simplifies to
 \begin{equation}
   \widetilde{\sigma}_{\alpha,\nu=1}(x)\,=\,
  \frac{1}{\pi^2}\,(x+\alpha)\,\frac{\cosh x/2}{\sinh(x+\alpha)/2}.
\end{equation}

The spectral theorem for the self-energy in frequency is
\begin{equation}
\Sigma(\omega) = - \frac{1}{\pi} \int_{-\infty}^\infty d\varepsilon \,\,
\frac{\Im\Sigma(\varepsilon)}{\omega + i 0^{+} - \varepsilon}.
\end{equation}
The scaling form corresponding to
\eqnref{eq:scaling_tau} is
\begin{equation}
\label{eq:sigma_scaling_omega}
    \Im\Sigma(\omega) = - T^\nu \sigma(\omega / T) = - \frac{\lambda
    \pi^{1+\nu}}{\cosh(\alpha/2)} T^\nu
    \widetilde{\sigma}_{\alpha,\nu}(\omega/T)
\end{equation}
where $\lambda$ is a non-universal prefactor and the normalization conventions
are chosen to match~\cite{parcollet_kondo_prb_1998}.

The real part of the self-energy also obeys scaling properties. Defining: 
\begin{equation}
    1-\frac{1}{Z(T,\omega)}\equiv \frac{1}{\omega}\left[\Re\Sigma(\omega,T)-\Re\Sigma(0,T)\right],
\end{equation}
we obtain
\begin{equation}
1-\frac{1}{Z} = \frac{1}{\pi}\,P\int d\varepsilon \frac{-\Im\Sigma(\varepsilon)}{\varepsilon(\omega-\varepsilon)}   
\end{equation}
For $\nu<1$, we can substitute the scaling form $-\Im\Sigma(\varepsilon) = T^\nu
\sigma(\varepsilon / T)$ in the integral without encountering divergences 
and obtain a universal contribution:
\begin{equation}
\frac{1}{Z(T,\omega=xT)}\,
=\,1-\frac{1}{\pi T^{1-\nu}} \,P\int dy \frac{\sigma(y)}{y(x-y)}
\end{equation}
For $\omega = 0$, at low-$T$:
\begin{equation}{}
Z(T,\omega=0)\,\simeq\,\pi T^{1-\nu} \left[\int dy \frac{\sigma^\prime(y)}{y}\right]^{-1}.
\end{equation}
For $\nu \geq 1$, the integral above diverges and $Z$ has no universal
contribution.

We can contrast the behavior of the skewed scaling functions, with the case of a
Fermi liquid. At low energy, the real-frequency self-energy has the form
\begin{equation}
\Im \Sigma_\mathrm{FL}(\omega) = - \frac{\omega^2 + (\pi T)^2}{E_\mathrm{FL}} - b\omega^3 -
c\omega T^2 + \cdots = - T^2 \sigma_{FL}(\omega/T).
\end{equation}
In the scaling regime, the higher subdominant corrections vanish and
$\sigma_\mathrm{FL}(x) = \pi^2 + x^2$ is
particle-hole symmetric.

\subsection{Extrapolating $\Im\Sigma$}
\label{app:extrap-imSigma}

Here, we will describe a procedure to extract $\Im\Sigma_\mathrm{extrap}(\omega
= 0)$, which is using in the lifetime (Fig.~\ref{fig:electron_sigma}) and the
transport scattering rate (Figs.~\ref{fig:app-im-sigma-extrap} and \ref{fig:app-scattering-rate-fl}). Since our
calculations are at finite temperature $T$, the self energy $\Sigma(\omega, T)$
is analytic even at the critical doping $\doping_c$. One can directly
extrapolate $\Sigma(i\omega = 0, T)$ from finite Matsubara frequencies,
although such a procedure always have ambiguities from the choice of fitting
function.

The analytic form of the self-energy $\widetilde{\sigma}(x)$ can guide our
choice of fitting function and improve our extrapolation estimate. In
particular, we can use the spectral representation
\begin{equation}
\Sigma(i\omega_n) = \frac{T^\nu}{\pi} \int_{-\infty}^\infty d\varepsilon \,\,
\frac{\sigma(\varepsilon/T)}{i\omega_n - \varepsilon}
\end{equation}
to can accurately fit the data at Matsubara frequencies. Then, we can then
obtain the zero frequency extrapolations by analytically continuing the known
expressions:
\begin{align}
\label{eq:sigma_scaling_zero}
\Im\Sigma(\omega=0^+) &= \frac{ T^\nu}{\pi} \Im\lim_{\delta\to
0^+} \int_{-\infty}^\infty d\varepsilon \,\,
\frac{\sigma(\varepsilon/T)}{0 + i\delta - \varepsilon} = - T^\nu 
\sigma(0)
\end{align}
The scaling ansatz of Eqs.~\eqref{eq:sigma_scaling_omega}
and~\eqref{eq:sigma_scaling_x} encompasses a variety of different behaviors,
including the marginal Fermi liquid case ($\nu = 1$) and the Fermi Liquid phase
($\nu = 2$). The procedure outline above, while an improvement over simple
extrapolation, is still approximate as it neglects sub-leading corrections to
scaling as well as non-universal UV corrections, present in the numerical data.

Equations~\eqref{eq:sigma_scaling_omega}
and~\eqref{eq:sigma_scaling_x} have three fitting parameters $\nu, \alpha,
\lambda$. Since $\Im\Sigma(i\omega_n)$ is only very weakly dependent on
$\alpha$ and therefore difficult to fit, we choose to use the estimate of
$\alpha$ from Fig.~\ref{fig:scaling_powers}b. We fit effective $\nu,
\lambda$ using the convenient scaling combination
\begin{equation}
\frac{\omega_0 \Im\Sigma(i\omega_n) - \omega_n \Im\Sigma(i\omega_0)}{\omega_n -
\omega_0} = \frac{ T^\nu \omega_n \omega_0 (\omega_n + \omega_0) }{\pi}
\int_{-\infty}^\infty d\varepsilon \,\,
\frac{\sigma(\varepsilon/T)}{(\omega_n^2 + \varepsilon^2)(\omega_0^2 +
\varepsilon^2)}
\end{equation}
over the range of Matsubara points $1 \leq n \leq 6$ for each value of doping
$\doping$ and inverse temperature $\beta$. We can then evaluate
$\Im\Sigma_\mathrm{extrap}(0)$ using \eqnref{eq:sigma_scaling_zero}. 
Finally, for this updated extrapolation, we do not consider high doping data $\doping >
0.6$, since they are weakly correlated and noise makes extracting
$\nu, \alpha, \lambda$ unreliable.

\emph{Note:} In a previous version of this manuscript, $\Im\Sigma_\mathrm
 {extrap}(0)$ was estimated using a polynomial spline. Although the spline
 result noticeably overestimated the value of $\Im\Sigma_\mathrm{extrap}
 (0)$, it mostly captured the correct temperature dependance. 

\subsection{Conductivity}
\label{app:conductivity}

In the main text, we considered an electronic model on a fully connected
lattice (long-range hopping and spin interactions). For this non-local model,
it is somewhat unnatural to define a conductivity. One way to address this is
to note that there are other lattice models which do have a natural conductivity
and that share exactly the same local self-energy $\Sigma(\omega)$
(identical EDMFT equations). For example, the non-disordered model in the DMFT
approximation, a translationally invariant lattice of fully connected `dots'
with internal random coupling, along the lines of e.g.
Refs.~\cite{balents2017,senthil2018}, or a high coordinate number
lattice~\cite{Georges1996}.



In all cases, we can define the conductivity via the Kubo formula:
\begin{equation}
    \sigma_{dc}= \pi e^2\int d\omega\, 
    \left(-\frac{\partial f}{\partial\omega}\right)\,
    \int d\vare\, \Phi(\vare)\, A(\omega,\vare)^2 
\end{equation}
In this expression, $\Phi(\vare)$ is the non-interacting transport function (density of states 
weighted by velocities) which depends on the choice of lattice and 
$A(\vare,\omega)=-\Im G(\omega+i0^+,\vare)$ is the spectral function. Vertex 
corrections are absent in this class of models with a local self-energy \cite{Georges1996}. 
When the scattering rate $2|\Im\Sigma|$ is small enough, this expression can be further 
simplified into:
\begin{equation}
    \sigma_{dc}= e^2\int d\omega\, 
    \left(-\frac{\partial f}{\partial\omega}\right)\,\frac{1}{2|\Im\Sigma|}\,
   \Phi(\omega+\mu-\Re\Sigma)
\end{equation}
Expanding at low-$T$ and changing the integration variable to $x=\omega/T$, 
the dominant term reads: 
\begin{equation}
    \sigma_{dc}= e^2\,\Phi(\vare_F)\,
    \int_{-\infty}^{+\infty} \frac{dx}{4\cosh^2(x/2)}\,\frac{1}{2|\Im\Sigma[T,\omega=xT)]|} 
\end{equation}
When the scattering rate obeys a scaling form: 
$-\Im\Sigma = T^\nu \sigma(\omega/T)$ with $\nu<1$, this yields: 
\begin{equation}
    \sigma_{dc}=\frac{A}{T^\nu}\, e^2\,\Phi(\vare_F)\,\,\,,\,\,
    A=\int_{-\infty}^{+\infty} \frac{dx}{4\cosh^2(x/2)}\,\frac{1}{2\sigma(x)} 
\end{equation}
As shown in Fig.~\ref{fig:app-im-sigma-extrap} below, the extrapolated value of 
$|\Im\Sigma(T,0)|$ is $T$-linear to a good approximation in the accessible 
range of $T$, hence corresponding to a resistivity which is $T$-linear to a good approximation. 

Importantly, $Z$ does not enter the expression of the conductivity. This is in
contrast to the inverse lifetime $1/\taustar$ (width of the spectral function)
as defined in the text:
\begin{equation}
    \frac{1}{\taustar}\,=\,\frac{1}{Z|\Im\Sigma(T,\omega=0)|}
\end{equation}
In this expression $Z$ stands for $Z(T,\omega=0)$ as defined in the previous
section. Since, as shown there, $Z\sim T^{1-\nu}$, we see that $1/\taustar$ is
expected to have $T$-linear Planckian behavior even if $\nu<1$. Indeed, we
observe that, although the extrapolated $|\Im\Sigma(T,\omega=0)|$ itself is
approximately $T$-linear (Fig.~\ref{fig:app-im-sigma-extrap}), $T$-linearity is
more accurately obeyed for $1/\taustar$.

It is actually important to note that the prefactor $c$ in 
\eqnref{eq:planckian} is of order unity only provided $Z$ is indeed included in
the definition of $\taustar$ - otherwise the prefactor is much larger. this is
in line with the remarks of Ref.~\cite{Sadovskii_2020}, and also with the
procedure used in the experimental
literature~\cite{Bruin_2013,Legros19,Grissonnanche_2020} in which the {\it
effective} mass is used to infer $1/\taustar$ from resistivity measurements.

\subsection{Entropy - free local moments}
\label{app:entropy}

It is instructive to note that in the atomic limit and for $U=\infty$ the entropy 
per site, for a fixed density $n$, is given by:
\begin{eqnarray}
    s(n)&=& - (1-n)\ln (1-n) -n \ln (n/2),\nonumber \\
    \frac{\partial s}{\partial n}&=&\ln \left(2\frac{1-n}{n}\right),\qquad
    \frac{\partial^2 s}{\partial n^2} = - \frac{1}{n(1-n)}
\end{eqnarray}
so that $s(n)$ has a maximum at $n=2/3$ (hole-doping $p=1/3$) and 
${\partial^2 s}/{\partial n^2}<0$ for all densities.

\subsection{Supplementary Data}
\label{app:supp-data}

Here we present some additional data to complement the data shown in the main
text.

\subsubsection{Fermionic Properties}
\begin{figure}[h] 
    \centering
    \includegraphics{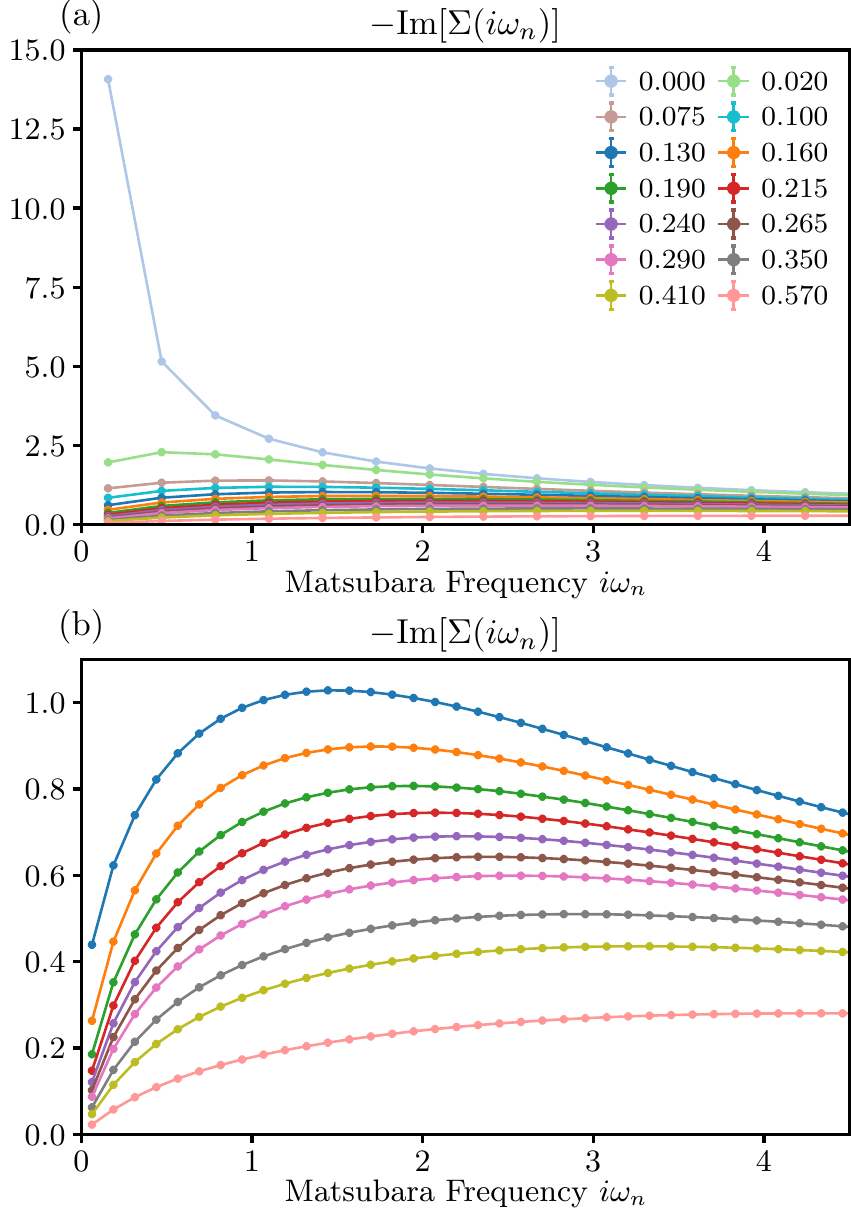}
    \caption{
    \label{fig:app-im-sigma-omega}  
    Imaginary component of the electronic self-energy $-\Im \Sigma(i\omega_n)$
    vs Matsubara frequency $i\omega_n$ at (a) $\beta = 20$  and (b) $\beta =
    50$. At half-filling $\delta=0$, the self-energy diverges at low
    frequencies $-\Im \Sigma(i\omega_n) \sim 1 / i\omega_n$, characteristic of
    the expected insulating behavior (see \cite{Cha2020}). At $\beta = 50$ and
    doping values $\doping = 0.130$-$0.190$ close to the critical point, the
    frequency dependance is more singular than linear close to $i\omega_n = 0$.
    This reflects the anomalous scaling discussion in
    Fig.~\ref{fig:scaling_powers} (main text).}
\end{figure}

\begin{figure}[h!] 
	\includegraphics{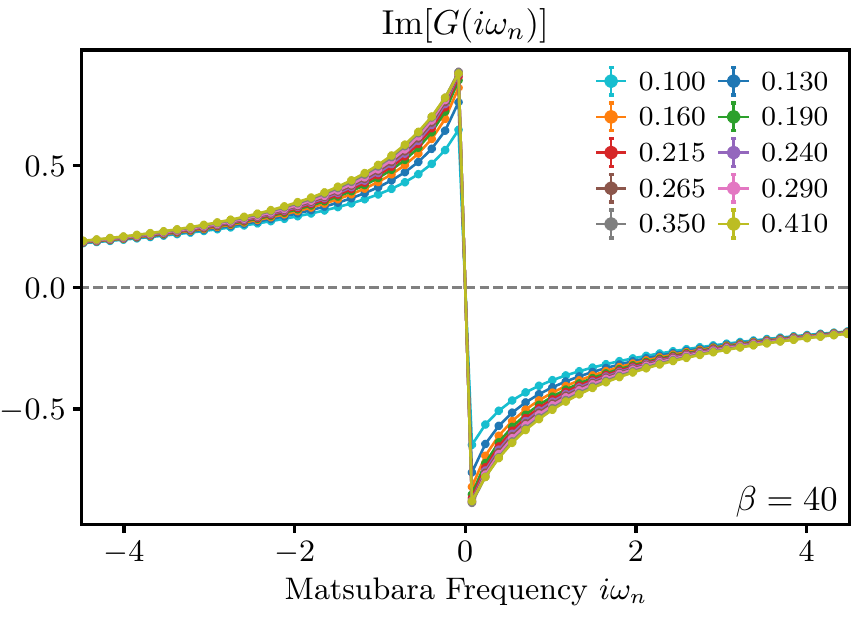}
	\caption{
	\label{fig:app-im-g-omega}  
     Imaginary component of the electronic Green function $\Im G(i\omega_n)$ vs
    Matsubara frequency $i\omega_n$ at $\beta = 40$. All doping values shown
    ($\doping > 0$) have a discontinuity at $i\omega_n = 0$, indicative of a
    metallic states. This is true both for doping in the spin-glass order
    regime $\doping \leq \doping_c$ where Luttinger's theorem breaks down as
    well as the Fermi liquid regime $\doping > \doping_c$. }
\end{figure}

\begin{figure}[h!] 
    \centering
    \includegraphics{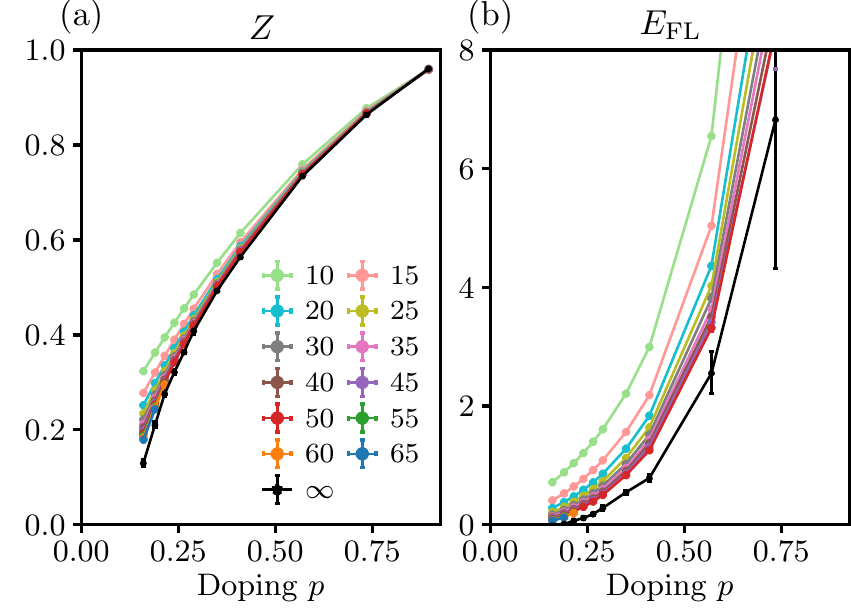}
    \caption{
    \label{fig:app-fl-fit}  
    Quasiparticle residue $Z$ and Fermi liquid coherence scale
     $E_{\mathrm{FL}}$ obtained by fitting $\Im\Sigma(i\omega_n)$ to the Fermi
     liquid form \eqnref{eq:sigma_fl}. For doping values close to the critical
     point, Fermi liquid theory does not apply, but we can extract $Z,
     E_{\mathrm{FL}}$ as effective coefficients of the linear and quadratic
     frequency terms.  For $\doping < \doping_c$ the self-energy has a finite
     zero-temperature intercept so the form \eqnref{eq:sigma_fl} does not
     apply.}
\end{figure}

\begin{figure}[h] 
    \centering
    \includegraphics{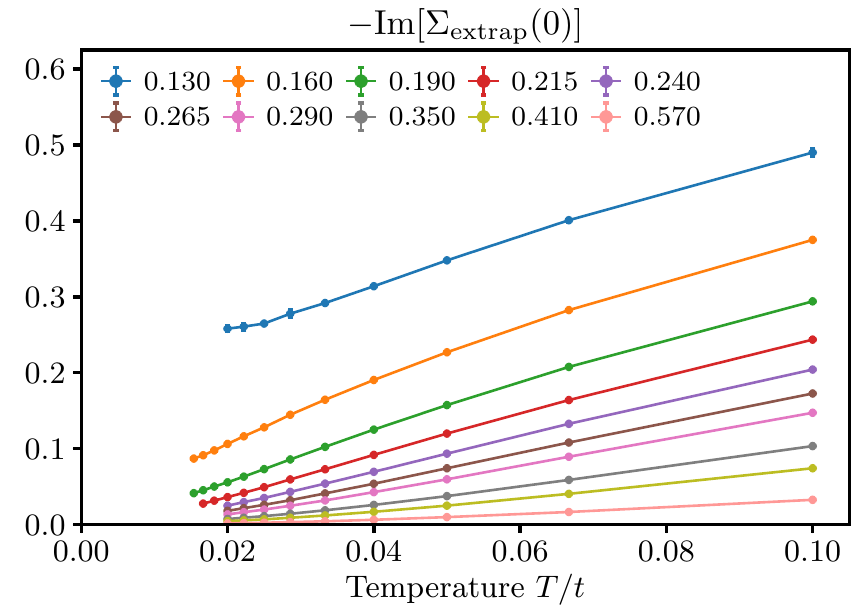}
    \caption{
    \label{fig:app-im-sigma-extrap}  
    Extrapolated value of the imaginary part of the self-energy. The quantity
    $-2 \Im\Sigma_\mathrm{extrap}(0)$ is the transport scattering rate.  It is
    interesting to note that it itself is approximately linear in temperature
    $T$ for doping near $\doping_c$ and low temperatures, which arises from the
    combination of effective finite temperature dependance of
    $\alpha,\nu,\lambda$ of \eqnref{eq:sigma_scaling_omega}. It is also used in
    determining the effective single-particle lifetime presented in
    Fig.~\ref{fig:electron_sigma}(c). For $\doping < \doping_c$, the data
    extrapolates to a finite value at zero temperature. }
\end{figure}

\begin{figure}[h] 
    \centering
    \includegraphics{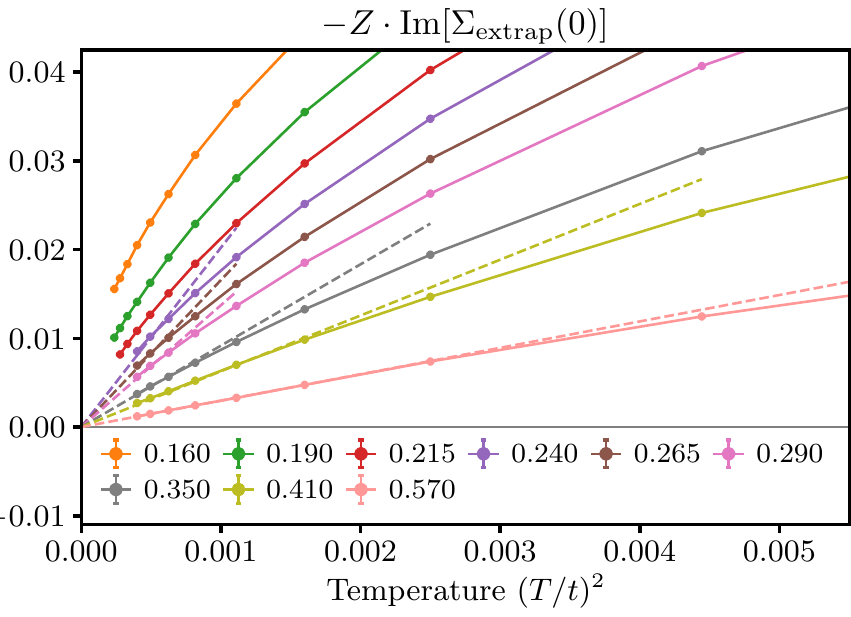}
    \caption{
    \label{fig:app-scattering-rate-fl}  
    Inverse effective single particle lifetime $- Z \Im[\Sigma_{\mathrm{extrap}}(0)]$.
    This is the same data as plotted in Fig.~\ref{fig:electron_sigma}(c), but
    focused on large values of doping.  The system follows the expected $\sim T^2$ Fermi
    liquid behavior (dashed lines) at large $\doping$ and temperatures below
    the cross-over scale $T_\mathrm{FL}$. }
\end{figure}

\begin{figure}[h] 
    \centering
    \includegraphics{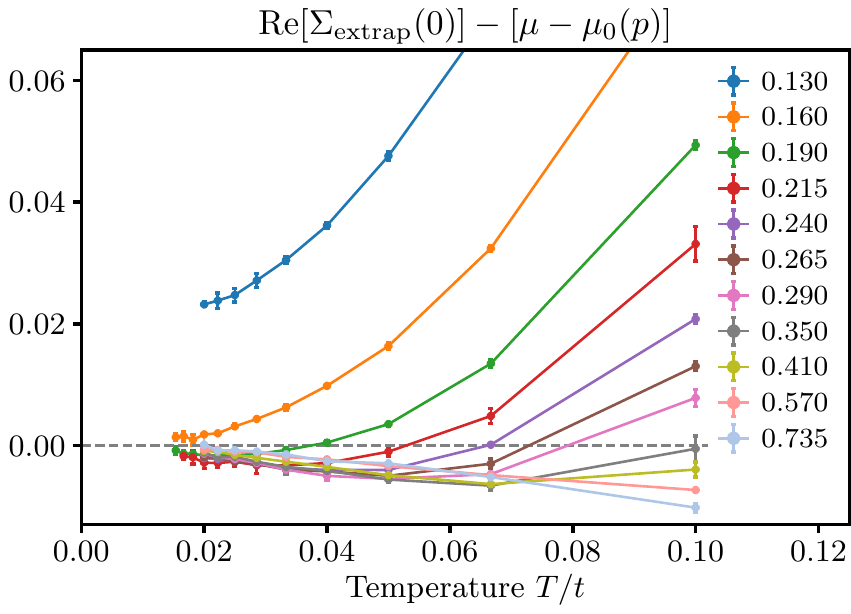}
    \caption{
    \label{fig:app-luttinger}  
    Finite temperature generalization of Luttinger's invariant $\Re
    \Sigma_{\mathrm{extrap}}(0) - [\mu - \mu_0(\doping)]$; this is the same
    data as in Fig.~\ref{fig:electron_sigma}b shown as a function of
    temperature $T$. At $\doping=0.130$, the data extrapolates to a large
    finite value which shows the transition to a phase distinct from the Fermi
    liquid at high doping.}
\end{figure}

Figures~\ref{fig:app-im-sigma-omega} and \ref{fig:app-im-g-omega} shows the
frequency dependance of the self-energy and Green functions. These show that
the state at half-filling is insulating, while at finite doping the system is
metallic on both sides of the critical point $\doping_c$. We also see signature
of the anomalous frequency scaling close to the critical doping. This a
complementary picture to the critical scaling shown in imaginary time in the
main text and also Fig.~\ref{fig:app-sigma-tau-scaling} below. In the Fermi liquid
regime we can fit the low frequency part of $\Im\Sigma(i\omega_n)$ to obtain
the Fermi liquid parameters $Z, E_\mathrm{FL}$, see Fig.~\ref{fig:app-fl-fit}.
Extrapolations of the finite frequency self energy to zero frequency are shown
for $ \Im\Sigma_\mathrm{extrap}(0)$ (Fig.~\ref{fig:app-im-sigma-extrap}), for
the effective single particle scattering rate $Z \Im\Sigma_\mathrm{extrap}(0)$
(Fig.~\ref{fig:app-scattering-rate-fl}) as well as Luttinger's parameter
$\Delta \vare_F(T) = \Re\Sigma_\mathrm{extrap}(0) - [\mu - \mu_0(\doping)]$
(Fig.~\ref{fig:app-luttinger}).


\begin{figure}[h] 
        \includegraphics{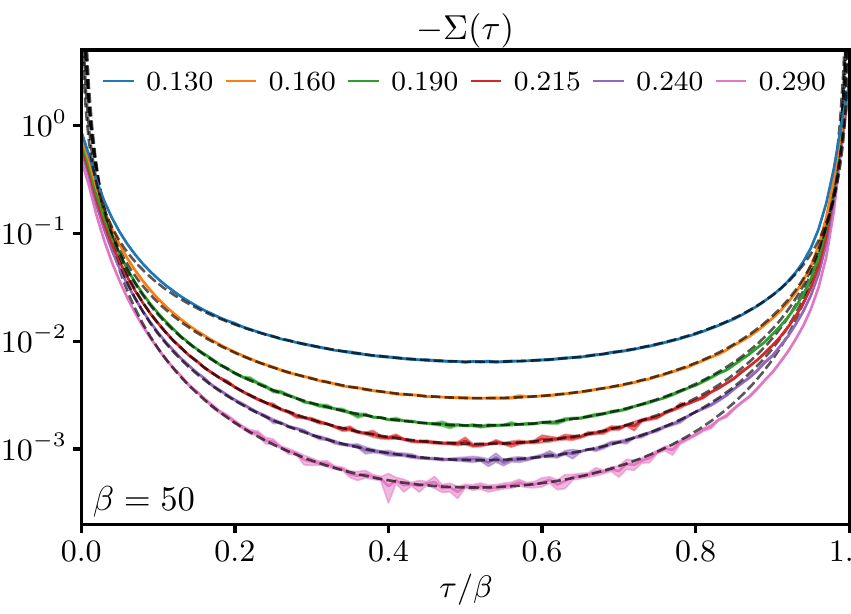}
    \caption{
    \label{fig:app-sigma-tau-scaling} Self-energy in imaginary time $-\Sigma(\tau)$ at low temperature and doping close the QCP. Dashed lines are expressions of the scaling function expected from conformal invariance [\eqnref{eq:sigmaScaling} of the main text], with $\alpha, \nu$ taken from fitting the data. This shows that the scaling form holds remarkably well for a large window of $\tau$ close to the long time limit $\tau = \beta / 2$.
        }
\end{figure}

\clearpage

\subsubsection{Bosonic Properties}

\begin{figure}[h] 
    \centering
    \includegraphics{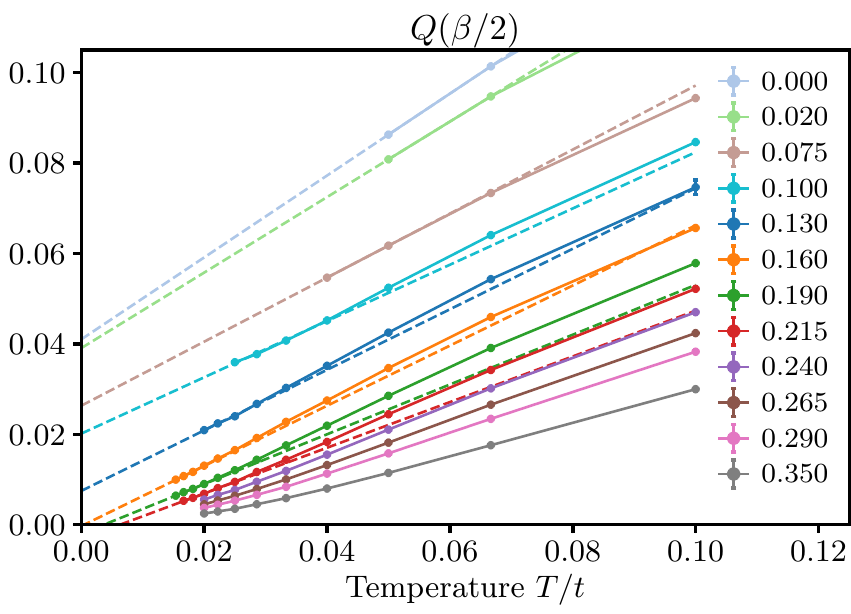}
    \caption{
    \label{fig:app-local-moment}  
    Long-time spin correlation $Q(\beta/2)$. For doping $\doping < \doping_c$,
    $Q(\beta/2)$ extrapolates to a finite value at zero temperature, indicating
    tendency of the paramagnetic solutions to form a local moment at low
    temperatures (cf.~Fig.~\ref{fig:chi}). }
\end{figure}

\begin{figure}[h] 
    \centering
    \includegraphics{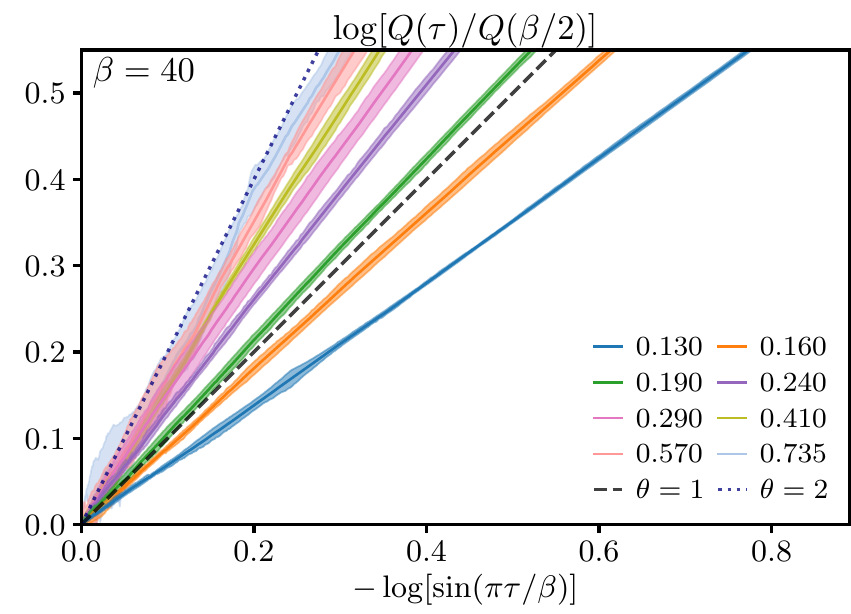}
    \caption{
    \label{fig:app-q-correlation-scaling}  
    Normalized spin-spin correlation $Q(\tau) / Q(\beta / 2)$ at fixed $\beta =
    40$. This shows the same data as in Fig.~\ref{fig:scaling_powers}(c),
    plotted so that the exponent $\theta$ is the slope of the data at small
    $-\log \vert\sin(\pi \tau/\beta)\vert$. Fermi liquid $\theta = 2$ (dotted)
    and SYK $\theta = 1$ (dashed) forms are shown. }
\end{figure}

\begin{figure}[h] 
    \centering
    \includegraphics{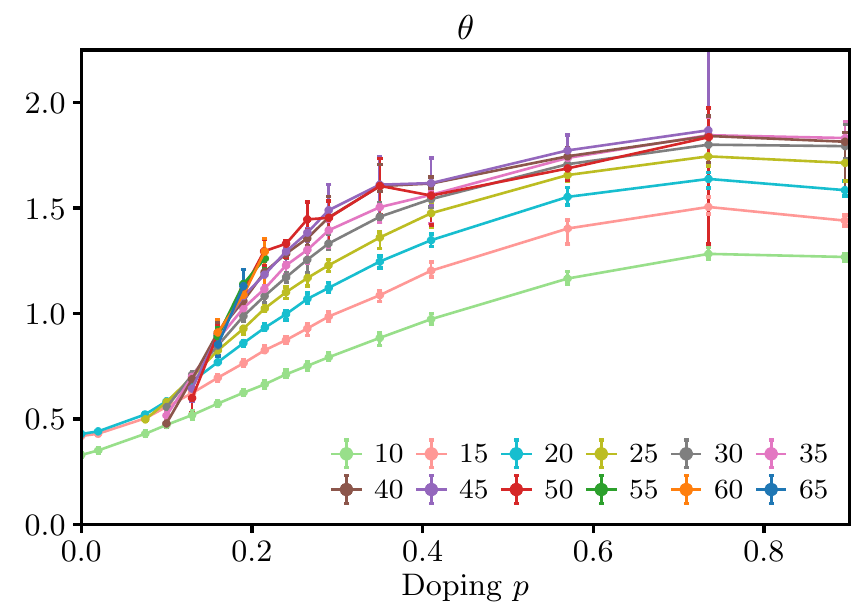}
    \caption{
    \label{fig:app-alpha-fit}  
    Power of the long-time spin correlation $Q(\tau)$, extracted from scaling
    plots such as Fig.~\ref{fig:app-q-correlation-scaling}, as a function of
    doping $\doping$ and for various $\beta$. This is the same data as is shown
    in the color-plot background in Fig.~\ref{fig:phase-diagram}. The curve
    $T_{\theta=1}$ of  Fig.~\ref{fig:phase-diagram}, is obtained by fitting the
    intersection of the interpolation of the current data with the constant
    line $\theta=1$; error propagation is done by bootstrap resampling. }
\end{figure}

Here we show some additional data related to the spin-spin correlation function
$Q$. Figure~\ref{fig:app-local-moment} shows the long time value of $Q$, and
shows how solutions with $\doping < \doping_c$ have tendency towards local
moment formations. Figures~\ref{fig:app-q-correlation-scaling}  and
\ref{fig:app-alpha-fit} are additional plots establishing the  critical scaling
form of $Q$ at long times.

\end{document}